\documentclass[twocolumn,prl,graphics,showpacs,noshowkeys]{revtex4}
\begin{document}
\input{epsf.sty}
\title{Intermittency is a consequence of 
turbulent transport 
in nonlinear systems 
}
\author{Benno Rumpf}
\affiliation{Max-Planck-Institute for the Physics of Complex Systems
N\"othnitzer Stra\ss e 38, 01187 Dresden, Germany}
\author{Alan C. Newell}
\affiliation{
Mathematics Department, University of Arizona, Tucson, AZ 85721, USA
}
\begin{abstract}
Intermittent high-amplitude structures emerge in 
a damped and driven discrete nonlinear Schr\"odinger equation 
whose solutions transport both energy and 
particles from sources to sinks. 
These coherent structures are necessary for any solution 
that has statistically 
stationary transport properties. 
\end{abstract}
\maketitle
Turbulent flows transferring energy from a stirring range at 
large scales to the dissipation range at small scales consist of 
dissimilar components, a broad spectrum of eddies 
and randomly occurring, intermittent coherent structures. The first 
cascade leads to Kolmogorov-like finite-flux power spectra. The 
second component is particularly 
visible in the anomalous short-scale behavior of the higher 
order moments of velocity differences. 
Compared to fully developed three dimensional turbulence, it 
is easier to quantify these two components in turbulent systems 
of weakly coupled dispersive waves. 
In this case, the Kolmogorov-Zakharov spectrum is a 
stationary finite flux solution 
of kinetic equations that follow from 
three and four wave resonances of the weak interactions. 
The second component emerges since the 
wave turbulence approximation \cite{nnb}  
is almost never valid at 
very low and very high wavenumbers where the 'weak' coupling approximation 
breaks down, leading to the emergence of fully nonlinear structures 
\cite{dya}. 
In short, despite the fact that the amplitudes are, 
on the average, small, the weakly nonlinear 
dynamics can lead to intermittent and localized high amplitude  
events and anomalies in high order moments. It may also lead to a
contamination of low order moments and to power spectra which, at least 
in some wavenumber ranges, are dominated by strongly nonlinear events. 
Nowhere is this more evident than in the illuminating 
studies of Cai, Majda, McLaughlin, and Tabak (CMMT) \cite{mmt} 
(later confirmed by 
Zakharov, et al. \cite{zak}) on damped, driven and on  
freely decaying weakly nonlinear dispersive one dimensional
wave systems. 
Indeed, CMMT found that, in damped and driven systems, there were some 
situations in which the nonlinear solutions dominated at almost 
all scales. In the freely decaying state, they found the Kolmogorov-Zakharov 
spectra were 
much more likely to appear. This of course was disappointing 
because the strength of the Kolmogorov-Zakharov solution of the undamped, 
undriven kinetic 
equation is that it describes exactly what would be expected if an energy 
source at $k=0$ feeds at a constant flux rate through an inertial 
range to a viscous sink at $k=\infty$. 
\\
The purpose of this Rapid Communication 
is to demonstrate in a simple but 
representative model that in driven, damped systems in which there are fluxes
of two conserved densities (energy and particle number), the realization of a
{\itshape statistically steady \/}  
state demands the existence of high amplitude 
coherent structures because they play a crucial role in balancing the 
energy flux budget. We show that a single cascade of weakly interacting 
waves transporting particles from a source to a sink leads to a steady 
net loss of energy of the system. Any steady solution requires contributions 
from the nonlinear terms in order to offset this energy loss. 
This is achieved by the intermittent formation and destruction 
of high-amplitude structures. 
Fig.1 shows the intermittent emergence of peaks for a 
\begin{figure}[b]
\epsfbox{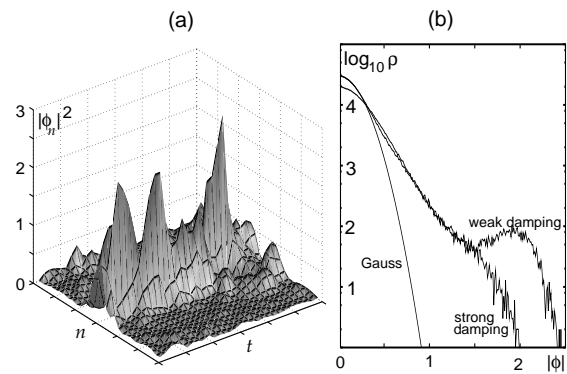}
\caption{
Numerical integration of the DNLS:   
(a) $|\phi_n|^2$ for a sector of 30 oscillators over 30 time units with 
strong damping.
(b) Density of sites with the amplitude $|\phi|$ as a function of 
$|\phi|$ 
for weak and for strong damping and a Gaussian fit of the density 
for weak damping. 
}
\end{figure}
discrete nonlinear Schr\"odinger equation 
\begin{equation}\label{nlsfd}
i\dot{\phi}_n+\phi_{n+1}+\phi_{n-1}-2\phi_{n}+
\phi_n^2\phi^*_n={\cal F} (\phi,t)-{\cal D} (\phi,t)
\end{equation}
for a chain of 
complex oscillators. $\cal D$ is a 
short-wavelength damping term while $\cal F$ drives the 
system on long space scales. 
Fig.1(a) displays the typical 
long-time behavior of $|\phi_n|^2$ in a sector 
of 30 lattice sites from a chain of $N=512$ oscillators 
with periodic boundary conditions over 30 time units. 
$|\phi_n|^2$ is small on average, but high-amplitude 
structures emerge locally. The peak shows a 'breathing' 
behavior, decreasing and increasing irregularly. 
Fig.1(b) shows the average 
density $\rho$ of oscillators with an amplitude $|\phi|$ 
for weak and for strong 
damping forces. The density for small $|\phi|$ is 
Gaussian, but for larger $|\phi|$ the intermittent high-amplitude 
structures lead to a slower non-Gaussian decay. 
In the weakly damped system, the density increases slightly for 
high amplitudes so that there is a small hump near $|\phi|=2$. 
The number of 
these intermittent events increases with the driving force, but the 
hump remains in the vicinity of 
$|\phi|=2$. \\
In the simulations we use time-periodic $\delta$-kicks as 
damping and driving forces 
$\cal D$ and $\cal F$, which allows a simple control of the 
energy input and output: 
The driving 
increases the homogeneous mode 
$c_0\rightarrow (1+\lambda_{\cal F})c_0$, and  
the damping decreases modes on a short space scale as 
$c_k\rightarrow (1-\lambda_{\cal D})c_k$ 
where $c_k=(1/N)\sum_n \phi_n exp(ikn)$ and the wavenumber $k$ is 
in the Brillouin-zone $[0,2\pi]$. 
In numerical studies we apply the damping to 
the short-wavelength modes with $7\pi/8\le k\le 9\pi/8$, 
so that the impact of the damping does not 
decrease with the system's size. Analytical studies are simplified 
by restricting the damping to the mode $k=\pi$. 
$\cal D$ and $\cal F$ are zero in the intervals of one time unit 
between the time-periodic kicks. 
No important changes are found for 
shorter intervals between the kicks, or for continuous 
driving and damping. 
\\
The dynamics between the kicks is governed 
by the nonlinear Schr\"odinger equation that 
derives as $i\dot\phi_n=\partial H/\partial \phi_n^*$ 
from the Hamiltonian 
\begin{equation}\label{hamilton}
\begin{array}{rcl}
H&=&H_2+H_4\\
&=&\sum_n 2\phi_n\phi^*_{n}-\phi_n\phi^*_{n+1}-\phi^*_n\phi_{n+1}\\
&-&\frac{1}{2}\sum_n\phi_n^2\phi^{*2}_n
\end{array}
\end{equation}
The modulus square norm or particle number 
\begin{equation}\label{particle}
A=\sum_n \phi_n\phi^*_{n}=N\sum_k c_kc_k^*
\end{equation}
is a second conserved quantity. 
In recent studies \cite{rune} we have shown that 
this isolated Hamiltonian system with no damping and driving force 
forms localized high-amplitude structures 
as a statistical consequence from the thermalization 
under the constraint of its two {\itshape conserved} quantities. 
Such peaks are generated in a self-focusing process of low-amplitude 
waves with long wavelengths and a low energy. 
A typical initial condition from which high-amplitude peaks emerge 
is the Rayleigh-Jeans distribution of the power 
$|c_k|^2=(\beta (\omega (k)+\gamma))^{-1}$ 
with a positive temperature $\beta^{-1}$. 
$\gamma$ is the chemical potential, $\omega(k)=2-2\cos k$ is the frequency. 
As the system approaches its state of maximum entropy, 
the spectrum of low-amplitude fluctuations becomes flat 
so that the power is equipartitioned on all modes ($\beta\rightarrow 0, 
\gamma\rightarrow \infty$). 
During this transformation of 
the low-energy spectrum to an equipartitioned spectrum,  
stable high-amplitude peaks with a negative energy emerge as a by-product 
of the production of entropy in the low-ampitude waves. 
No such peaks emerge from low-amplitude short waves 
with a high energy. 
\begin{figure}[t]
\epsfbox{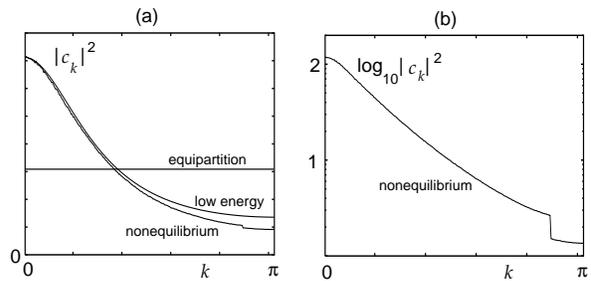}
\caption{
(a): Spectrum for the weakly damped nonequilibrium 
system compared to a similar low-energy thermodynamic 
spectrum $(\beta (\omega (k)+\gamma))^{-1}$ of the corresponding undamped 
and undriven Hamiltonian system. (b): Logarithmic plot 
of the spectrum for strong damping. The steps of the spectra 
at $k=7\pi/8$ result from the damping. 
}
\end{figure}
\\ 
In contrast, the long time behavior of solutions of the damped and driven 
system is governed, not by the values of $H$ and $A$, but by the 
{\itshape fluxes \/} of both quantities. 
Each driving step ${\cal F}$ changes the amplitudes $\phi_n$ by 
$\Delta \phi_n =\lambda_{\cal F} c_0$, so that 
the total number of particles increases by 
$
\Delta_{\cal F} A=\sum_n \phi_n\Delta\phi_n^*+\Delta\phi_n\phi_n^*=
2\lambda_{\cal F} N |c_0|^2$ 
while the damping changes the particle number by 
$
\Delta_{\cal D} A=-2\lambda_{\cal D} N |c_{\pi}|^2
$ 
for small values of $\lambda_{{\cal F},{\cal D}}$. 
Consequently, there is a flow of particles from 
the source at $k=0$ to the sink at $k=\pi$, and the power  
spectrum (Fig.2(a)) decays with the wavenumber. 
With most power gathered at low wavenumbers, the spectrum is 
similar to a thermodynamic spectrum (Fig.2(a)) 
of the corresponding undamped and undriven Hamiltonian system with 
a low energy. The high-amplitude structures are generated just by 
the same mechanism in such an isolated Hamiltonian system and in 
the damped and driven system where the permanent particle flow maintains 
the bias of the spectrum. 
The spectrum decays even exponentially for strong damping forces Fig.2(b). 
Particle loss and gain are balanced when the flow 
in a sufficiently large system is constant so that 
\begin{equation}\label{partbal}
\Delta_{\cal F}A +\Delta_{\cal D}A=0
\end{equation}
or $\lambda_{\cal F}|c_0|^2= \lambda_{\cal D} |c_{\pi}|^2$. 
The driving parameter $\lambda_{\cal F}$ and the damping parameter 
$\lambda_{\cal D}$ regulate the particle number and the particle flux 
of the system. By choosing the driving parameter as 
$\lambda_{\cal F}=\frac{\Delta A}{2N|c_0|^2}$
we obtain a constant flow of particles 
$\Delta_{\cal F}A=\Delta A$ into the system in the numerical simulations. 
The damping parameter $\lambda_{\cal D}$ is fixed. 
For a strong damping to forcing ratio, 
$\lambda_{\cal D}=\epsilon^{-2} \lambda_{\cal F}$ with 
$\epsilon^2\ll1$,  
it follows from (\ref{partbal}) that $|c_{\pi}|=\epsilon |c_0|$ and 
\begin{equation}\label{cocpi}
\lambda_{\cal F}|c_0|=\epsilon \lambda_{\cal D} |c_{\pi}|
\end{equation}
The input and output of particles also 
changes the quadratic coupling energy and 
the quartic energy. Gains and losses again have to match in a 
stationary nonequilibrium state so that 
\begin{equation}\label{hbal}
\Delta_{\cal F} H_2+\Delta_{\cal F} H_4+
\Delta_{\cal D} H_2+\Delta_{\cal D} H_4=0
\end{equation}
In order to understand the role of the strongly nonlinear terms 
in the energy flow, we analyze 
how it divides up among these four terms. 
The change of the quadratic coupling energy 
$H_2=N\sum \omega(k)c_k c_k^*$ is given by 
the flux of particles times the frequency $\omega(k)$. 
The influx of particles through the driving force 
leads to a zero energy influx 
$
\Delta_{\cal F} H_2=\omega(0) \Delta A=0
$
since $\omega(0)=0$.
The damping leads to a loss of coupling energy 
$
\Delta_{\cal D} H_2=-\omega(\pi) \Delta A
$
with $\omega(\pi)=4$. 
Consequently, the particle flux leads to a net loss of 
coupling energy 
$\Delta H_2=
\Delta_{\cal F} H_2 +\Delta_{\cal D} H_2 = -4\Delta A$. 
\begin{figure}[t]
\epsfbox{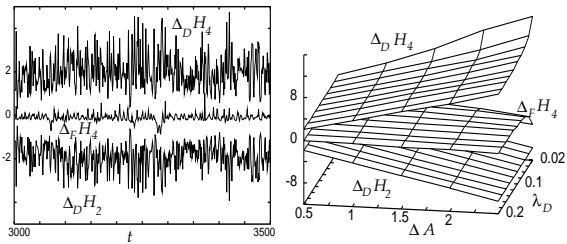}
\caption{
(a): Change of $H_2$ due to the damping 
and $H_4$ due to the damping and driving force as a function of time. 
(b): Time average of the same energy flows as a function of the 
particle flow $\Delta A$ and the damping constant $\lambda_{\cal D}$. 
}
\end{figure}
This energy loss must be 
balanced by an energy gain in the nonlinear term $H_4$ 
for maintaining a stationary nonequilibrium state. 
The viscosity changes the nonlinear energy 
$H_4=-(1/2)\sum \phi_n^2{\phi_n^*}^2$ 
as 
\[
\Delta_{\cal D} H_4= 2\lambda_{\cal D} 
N  Re(d_{\pi}c_{\pi}^*)
\]
where $d_k=(1/N)\sum_n |\phi_n|^2\phi_n e^{ikn}$. 
The driving changes this energy contribution 
by 
\[
\Delta_{\cal F} H_4= -2\lambda_{\cal F} 
N  Re(d_{0}c_{0}^*)
\]
Fig.3(a) shows 
$\Delta_{\cal D} H_2$, $\Delta_{\cal F} H_4$ and $\Delta_{\cal D} H_4$ as 
functions of time for a chain of 512 oscillators 
for a total particle flux per time unit $\Delta A=0.512$ 
and a strong damping to forcing ratio where $\lambda_{\cal D}=0.2$. 
This corresponds 
to the simulation of Fig.1(a), the strong damping case of 
Fig.1(b), Fig.2(b) and to Figs.4(c),(d). 
The fluctuations indicate single collapse 
events which have a smaller relative strength for larger system sizes. 
Fig.3(b) shows the time average 
of $\Delta_{\cal D} H_2$, 
$\Delta_{\cal F} H_4$, $\Delta_{\cal D} H_4$ as functions of 
the damping coefficient $\lambda_{\cal D}$ and the particle flux $\Delta A$, 
where the weakly driven case ($\Delta A=0.512$) corresponds to all other 
simulations. 
The impact of the driving force on the energy 
$\Delta_{\cal F} H_4\approx 0$ is negligible 
compared to the two effects of the damping $\Delta_{\cal D} H_2$ 
and $\Delta_{\cal D} H_4$. 
This predominance of 
$\Delta_{\cal D} H_4$ over $\Delta_{\cal F} H_4$ 
(i.e. 
$\lambda_{\cal F}Re(d_{0}c_{0}^*)\ll \lambda_{\cal D}Re(d_{\pi}c_{\pi}^*)$
) 
follows from $\lambda_{\cal F} |c_{0}|\ll \lambda_{\cal D}|c_{\pi}|$ 
by equation (\ref{cocpi}) 
and from $|d_{\pi}|\sim {\cal O}(|d_0|)$. The second relation is due 
to the spiky shape of $|\phi_n|^2\phi_n$ that leads to a flat power spectrum. 
In addition, the phases of 
$d_{\pi}$ and $c_{\pi}$ are correlated by the phase velocity of the peaks 
while $d_0 c_0^*$ oscillates randomly as it is governed by low-amplitude 
fluctuations. 
$\Delta_{\cal F} H_4$ 
yields a significant negative contribution only for high particle 
fluxes $\Delta A$ and weak damping forces $\lambda_{\cal D}$ 
(at $\lambda_{\cal D}=0.02$, $\Delta A=2.5$ in Fig.3b). \\
For all other cases, both $\Delta_{\cal F} H_2$ and 
$\Delta_{\cal F} H_4$ are zero or close to zero, 
so that (\ref{hbal}) reduces to 
$\Delta_{\cal D} H_2\approx -\Delta_{\cal D} H_4$. 
This yields  
$
Re(c^*_{\pi}d_{\pi})\approx 4c_{\pi}^*c_{\pi}
$
or the inequality 
\begin{equation}\label{dpicpi}
|d_{\pi}|\ge 4|c_{\pi}|
\end{equation}
\\
The importance of the nonlinearity $\Delta_{\cal D} H_4$ 
in the energy balance is reflected in the system's first three moments: 
Defining $\psi_n=\phi_n (-1)^n$, the inequality 
(\ref{dpicpi})  reads  
$|\sum |\psi_n|^2\psi_n| \ge 4|\sum \psi_n|$. 
The first moment $|\sum \psi_n|\sim \sqrt{\Delta A}$ 
increases with the particle flux. The second moment 
$\sum |\psi_n|^2= A$ is the particle number, which is small 
if the system is strongly damped and weakly driven. 
Equation (\ref{dpicpi}) shows that the third moment exceeds the first moment, 
so that the distribution is strongly non-Gaussian. 
High-amplitude excitations (Fig.1(a)) leading to 
the non-Gaussian tails of Fig.1(b) are necessary 
to maintain a stationary nonequilibrium state under the flow of 
energy and particles. \\
To understand how the high-amplitude structures balance the energy flow, 
we compute the input and output of energy and particles as 
functions of the amplitude $|\phi|$ of the oscillators. 
The driving kick feeds a number 
$\Delta_{\cal F}a(|\phi|) d|\phi|
=\sum_{n: |\phi|\le|\phi_n|\le |\phi|+d|\phi|} 
2Re(\phi_n \Delta_{\cal F} \phi_n^*)$ 
of particles to those lattice sites where where the amplitude 
is between $|\phi|$ and $|\phi|+d|\phi|$. 
The total particle gain as used in (\ref{partbal}) is 
$\Delta_{\cal F}A=\int_0^{\infty} \Delta_{\cal F}a(|\phi|)d|\phi|$. 
Similarly, the forcing changes the nonlinear energy at these 
lattice sites by 
$\Delta_{\cal F}h_4(|\phi|)d|\phi|=
\sum_{n: |\phi|\le|\phi_n|\le |\phi|+d|\phi|} 2|\phi_n|^2 Re(\phi_n^*\Delta_{\cal F} \phi_n)$. 
The increments of the damping $\Delta_{\cal D} a(|\phi|)$ and 
$\Delta_{\cal D}h_4(|\phi|)$ are defined analogously. 
$\Delta_{\cal F}h_2(|\phi|)$ is again zero, and 
$\Delta_{\cal D}h_2(|\phi|)=\omega(\pi)\Delta_{\cal D}a(|\phi|)$. 
Fig.4 shows the time average of these 
influxes and outfluxes of particles and energy 
as functions of $|\phi|$. 
In the weakly damped case (Fig.4(a),(b) with 
$\lambda_{\cal D}=0.02$, $\Delta A=0.512$), 
the particles are fed into the system mainly at lattice sites with 
moderate amplitudes $|\phi|\approx 0.5$ (Fig.4(a)), 
while particles are removed mainly at sites with high 
amplitudes $|\phi|\approx 2$. While the input of particles has a negligible 
effect on the balance of nonlinear energy, the output of particles 
leads to a gain of energy peaked at $|\phi|\approx 2$. 
Remarkably, a very small number of sites with high amplitudes 
is responsible for practically the total outflux of particles and 
the corresponding feedback of nonlinear energy. \\
This energy gain can be seen for the most simple case of 
an isolated real peak $\phi_n=\chi \delta_{nl}$ whose height 
decreases by $\Delta_{\cal D}\phi= \lambda_{\cal D} c_{\pi}\ll\chi$ 
during the damping kick, so that the 
coupling energy changes by 
$\Delta_{\cal D} H_2=-8\chi\Delta_{\cal D}\phi$. 
This increases the nonlinear energy by 
$\Delta_{\cal D} H_4=2\chi^3\Delta_{\cal D}\phi$. 
These fluxes are balanced if the peak amplitude is 
$\chi=2$ which agrees with the results of Figs.4(a),(b). 
\begin{figure}[t]
\epsfbox{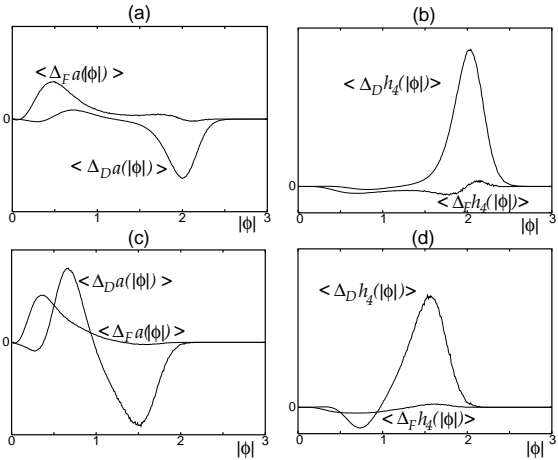}
\caption{
Average loss and gain of particles (a),(c) and of nonlinear energy (b),(d) 
as functions of the amplitude $|\phi|$. (a),(b) is the weak damping 
case, (c),(d) is the strong damping case. 
}
\end{figure}
Particles removed from the tip of the peak 
have to be replaced constantly through the driving mode at $k=0$. 
During the Hamiltonian evolution, 
the focusing process increases the peak again while short-wave 
fluctuations are radiated. 
\\
The input of particles in the strongly damped equation 
(Fig.4(a),(b) with $\lambda_{\cal D}=0.2$, $\Delta A=0.512$)
is maximal for 
low amplitudes $|\phi|\approx 0.4$ similarly to the weakly damped case 
of Fig.4(a). The damping leads to a significant loss of particles 
at $|\phi|\approx 1.5$, and to a tiny loss 
of particles in the domain of low-amplitude 
fluctuations at $|\phi|\approx 0.2$, but, 
paradoxically, also to a particle gain at 
$|\phi|\approx 0.7$. 
While there is still a net loss of particles, 
this partial recycling of particles increases 
the ratio of energy gain per particle loss. 
This allows the system 
to satisfy the balance condition (\ref{dpicpi})
with a peak amplitude less than $|\phi|=2$. 
As a simple model for this mechanism we assume two peaks at the sites 
$l$ and $l+3$ where 
$\phi_l=\chi>\xi=\phi_{l+3}$ are real and positive. 
A damping kick decreases the higher peak to 
$\phi_l=\chi-\Delta_{\cal D}\phi$ and increases the lower peak to 
$\phi_{l+3}=\xi+\Delta_{\cal D}\phi$. The decrease of the higher peak 
$\phi_l$  models the particle loss at $|\phi|\approx 1.5$ in Fig.4(c) 
and increase of the lower peak $\phi_{l+3}$ represents the gain of particles 
at $|\phi|\approx 0.7$. 
The coupling energy loss is 
$\Delta_{\cal D} H_2=-8(\chi-\xi)\Delta_{\cal D}\phi$ 
and the nonlinear energy gain is 
$\Delta_{\cal D} H_4=2(\chi^3-\xi^3)\Delta_{\cal D}\phi$. 
The gain of nonlinear energy can 
outweigh the loss of coupling energy for $\chi>2/\sqrt{3}$. 
\\
In summary, 
coherent structures are an essential component of the transport of 
particles from the large scales to small dissipation scales. 
The particle flow leads to a steady loss of coupling energy, 
so that the fluctuations have a low ratio of energy per particle. 
Such long-wave fluctuations form high-amplitude structures \cite{rune}, 
and it is the role of the coherent structures to 
balance the energy input and output by exploiting 
the nonlinear component of the energy. 
To offset this energy loss, high amplitude structures 
containing a significant amount of negative nonlinear energy must be formed 
and destroyed. Energy is transfered from the nonlinearity $H_4$ to 
the coupling $H_2$ during the peak growth, and 
pruning or destroying the peaks increases 
the systems total energy. The peaks 
are essentially the pipes through which energy and particles flow 
under the constraints of the balance conditions. \\
Obviously, this effect follows purely from the 
dispersion and the nonlinearity, and it is not restricted 
to the one dimensional discrete system of our simulations. 
A similar behavior was found \cite{dya} for the two-dimensional 
continuous focusing nonlinear Schr\"odinger equation when the 
forcing is applied at a midscale
wavenumber and the damping at large wavenumbers. 
This leads to an energy flux to high wavenumbers, and to an 
inverse particle flux to low wavenumbers. The particle flux 
builds condensates from which collapsing solitons emerge 
that carry particles to the dissipation scale and feed both 
particles and energy back into the wave field at these 
high wavenumbers.  
With an almost Gaussian statistics of the waves 
and a Poissonian distribution of the intermittent coherent events, 
the system resembles a two species gas. 
The presence of collapses does not 
appear to affect low moment statistics of two- or three-dimensional systems 
if the damping is sufficiently strong, while the collapses 
contaminate the power spectrum generated by 
wave-wave interactions in nonintegrable one dimensional systems. 
The important point to stress is that no particles would be dissipated and  
no steady state would be achieved without strongly nonlinear collapses.

\end{document}